# Diagnosis of COVID-19 and Non-COVID-19 Patients by Classifying Only a Single Cough Sound

**Masoud MALEKI (Mesut MELEK)** (Asst. Prof. Dr.)

Gumushane University

Department of Electronics and Automation

29100 Gumushane/TURKEY

Phone (mobile): (+90) 538 578 7035

E-mail: masoud.maleki1361@gmail.com

E-mail: mesutmelek@gumushane.edu.tr

ORCID: 0000-0002-7152-7788

# Diagnosis of COVID-19 and Non-COVID-19 Patients by Classifying Only a Single Cough Sound

Masoud MALEKI

*Department of Electronics and Automation, Gumushane University, Gumushane, Turkey*

**Abstract**

In the last month of 2019, a new virus emerged in China, spreading rapidly and affecting the whole world. This virus, which is called corona, is the most contagious type of virus that humanity has ever encountered. The virus has caused a huge crisis worldwide as it leads to severe infections and eventually death in humans. On March 11, 2020, it was announced by the World Health Organization that a COVID-19 outbreak has occurred. Computer-aided digital technologies, which eliminate many problems and provide convenience in people's lives, did not leave humanity alone in this regard and rushed to provide a solution for this unfortunate event. One of the important aspects in which computer-aided digital technologies can be effective is the diagnosis of the disease. Reverse transcription-polymerase chain reaction (RT-PCR), which is a standard and precise technique for diagnosing the disease, is an expensive and time-consuming method. Moreover, its availability is not the same all over the world. For this reason, it can be very attractive and important to distinguish the COVID-19 disease from a cold or flu through a cough sound analysis via smartphones which have entered into the lives of many people in recent years. In this study, we proposed a machine learning-based system to distinguish patients with COVID-19 from non-COVID-19 patients by analyzing only a single cough sound. Two different data sets were used, one accessible for the public and the other available on request. After combining the data sets, the features were obtained from the cough sounds using the mel-frequency cepstral coefficients (MFCCs) method, and then they were classified with seven different machine learning classifiers. To determine the optimum values of hyperparameters for MFCCs and classifiers, the leave-one-out cross-validation (LOO-CV) strategy was implemented. Based on the results, the k-nearest neighbors classifier based on the Euclidean distance (k-NN Euclidean) with the accuracy rate, sensitivity of COVID-19, sensitivity of non-COVID-19, F-measure, and area under the ROC curve (AUC) of 0.9833, 1.0000, 0.9720, 0.9799, and 0.9860, respectively, is more successful than other classifiers. Finally, the best and most effective features were determined for each classifier using the sequential forward selection (SFS) method. According to the results, the proposed system is excellent compared with similar studies in the literature and can be easily used in smartphones and facilitate the diagnosis of COVID-19 patients. In addition, since the used data set includes reflex and unconscious coughs, the results showed that conscious or unconscious coughing has no effect on the diagnosis of COVID-19 patients based on the cough sound.

**1. Introduction**

A pandemic was declared by the World Health Organization on March 11, 2020. The cause of the disease was stated to be the new coronavirus 2 (SARS-CoV-2), which causes severe acute respiratory syndrome [1]. This epidemic disease, called COVID-19, affected the lifestyle, economy, social life, and education of billions of people. This disease, which is highly contagious and has no fully medically proven cure, has caused more than 1.94 M deaths worldwide by January 2021. The symptoms in patients with COVID-19 vary significantly depending on the individual, and it may take up to 14 days for the symptoms to appear [2]. Fever, fatigue, and dry cough are the most common symptoms [3] which can easily be mistaken for a cold or flu [2].

Since the day the pandemic started, healthcare teams around the world have been working on the diagnosis, follow-up, and treatment of patients. Moreover, most of the researchers are trying to help humans to go through these difficult days more easily by examining this event in their field [4], [5], [6], [7]. In this case, studies on COVID-19 are expected from artificial intelligence (AI) and machine/deep learning (ML) techniques. These techniques, which are generally known as computer-aided digital technologies, have affected and changed human life, especially in recent years. A system based on machine learning techniques is an intelligent system that gains experience from past occurrences and adapts to new situations without the need for explicit programming [8], [9]. These systems are used to solve a variety of computer science problems, from bio-informatics to image processing [10]. Therefore, machine learning systems and computer-aided digital technologies, in general, can be used on many fronts to combat COVID-19 [11], [12].

Early diagnosis of COVID-19 is important as many other diseases. The standard and definitive diagnosis of the COVID-19 is made via the reverse transcription-polymerase chain reaction (RT-PCR) test of the infected secretions from the nasal or throat cavity [13]. The results of this test can sometimes take up to 48 hours to come out. In addition, in order for the test to be effective, patients must remain isolated during this time. RT-PCR test is not only time-consuming but also expensive and problems occur in large-scale deployments [14]. The cost of each test in the United States is approximately 23 dollars [15]. The governments try to test as much as they can every day, as they do not have the chance to test and control the whole country on one day. Moreover, the availability of the RT-PCR test is not the same all over the world. Therefore, it is of great importance to have a fast, simple, accurate, cheap, and easily accessible test.

One of the aspects in which machine learning can be effective is the early diagnosis of COVID-19. As mentioned previously, one of the most common and early symptoms of COVID-19 is coughing [2]. If non-coronavirus-induced coughs are distinguished from coronavirus-induced coughs through machine learning techniques, a cost-effective, easy, fast, and early diagnosis system can be offered. In this way, in addition to the low cost, suspected candidates can record the cough sounds on their smartphones whenever they want and perform a preliminary determination for their status. It can also reduce the burden of healthcare teams by effectively reducing the congestion in hospitals. This system, which is based on cough sounds, can also be used as a scanning method in airports, buses, waiting rooms of hospitals, nursing homes, and similar crowded environments [16].

In humans, different viruses, bacteria, or other acute and chronic health conditions, or even substances such as smoke and dust entering the lungs can cause coughing. In medicine, it is important for physicians to know whether the cough is wet, dry, or a wheezing, and whooping cough, in addition to how often and how severely the patient coughs [16]. Machine learning-based systems can detect the type of coughs (wet, dry, wheezing, and whooping cough) by providing medical professionals with more accurate clinical information about the frequency and severity of cough episodes. For this reason, even before the COVID-19 outbreak, studies on these issues were implemented. In a study [17], which was performed in 2011, two features that can be used to analyze cough sounds and distinguish between dry and wet cough sounds were identified. However, a clear distinction was observed by using only eight dry and eight wet cough sounds.

Pramono et al. [18] presented a system for diagnosing whooping cough in young children, which can be fatal if untreated. In their study, audio recordings from 38 patients were used for automatic diagnosis of pertussis by analyzing the cough and whooping sounds. The algorithm was able to successfully detect whooping cough from sound recordings and automatically detect individual cough sounds with 92% accuracy. As mentioned earlier, the frequency of coughs provides information for physicians to diagnose diseases. In [19], deep neural networks were used for cough detection. The accuracy rate of the system was reported to be 82.5% for the three classes defined, namely, cough, speech, and other. A preprocessing method was proposed for the detection of coughs in a noisy environment [20]. Next, a methodology for automated analysis of cough sounds using support vector machines (SVM) was presented.

Since the start of the pandemic, researchers working on computer-aided digital technologies have offered different ideas, solutions, and methods based on previous experiences. This covers a range from the analysis of the CT scans and X-ray images [21], [22], [23], [24], [25] for the diagnosis of COVID-19 to emotional and sentiment analysis from social media [26], [27], [28]. One of the parameters that has the greatest impact on machine learning studies is the data set. Since COVID-19 is a newly emerging disease and more importantly considering the status of the COVID-19 patients, it is very difficult to collect and access data sets. However, despite all these difficulties, studies on sound and especially on cough sound appear in the literature. Although most of these studies are not peer-reviewed yet, they can be obtained from different preprint banks. In [29], respiratory sounds of COVID-19 patients, with the help of a binary classifier, were distinguished from respiratory sounds of healthy people with an area under the curve (AUC) exceeding 0.80. Ali et al. [30] presented a mobile application that records and analyzes 3-second cough sounds through an application called AI4COVID-19. A total of 328 cough sounds of four different types including COVID-19, asthma, bronchitis, and healthy from 150 people were recorded and classified. The accuracy rate of the system was calculated as 92.85%. In a study based on deep neural network (DNN), coughs of people with COVID-19 were distinguished from those of healthy people with an accuracy of 96.83% [31]. In [32], cough sounds that were collected from 3621 people via mobile phones were classified with 0.72 AUC. In [33] cough sounds were classified with a 95.86% accuracy rate with SVM's RBF kernel function classifier by obtaining features by the MFCCs method. The sensitivity of the system to COVID-19 cough sounds was calculated as 98.6% and the sensitivity was obtained as 91.7%.

Most of the studies conducted have been based on recordings involving a few coughs. That is, for example, a 9-second recording has 3 or 4 coughs. In addition, all of the studies have been conducted on mandatory and conscious cough sounds. In this study, considering these two important points, two different datasets were combined and used. The virufy [34] data set is a public data set and contains 121 single cough records. The novel coronavirus cough database (NoCoCoDa) [16] is available to researchers free of charge upon request. This data set contains 73 single coughs that include reflex COVID-19 cough sounds and are not mandatory. The features were obtained by the MFCCs method on the data set acquired by combining these two data sets and classified with seven different classifiers. The optimum values of the hyperparameters of the system were determined based on the leave-one-out cross-validation (LOO-CV) strategy. Finally, effective features were determined separately for each classifier using the sequential forward selection (SFS) method. In this way, mandatory and conscious single cough sounds, in addition to reflex single cough sounds of COVID-19 patients, were successfully distinguished from single cough sounds of non-COVID-19 patients in the present study. The results showed that the proposed system is more successful than other systems. In addition, the results revealed that conscious or unconscious coughs have no effect on the diagnosis of COVID-19 patients with cough sounds.

In the following section, materials and methods are explained. In the third section, the results are given and in the fourth section, the results are discussed. Section 5 presents the conclusion.

## 2. Materials and methods

### 2.1 Data set description

#### 2.1.1 Virufy COVID-19 open cough data set

The virufy COVID-19 open cough data set is the first free, publicly available data set containing COVID-19 cough sounds [34]. COVID-19 PCR test results were also given along with the demographics of all the patients in the data set. After obtaining informed patient consent, the data were collected from patients in a hospital and under surveillance and verified by physicians, following standard operating procedures (SOPs). These cough sounds, which were collected from 16 patients, were recorded at a sampling frequency of 48 kHz. Then each recording was split so that it contained only one cough with the duration of 1.645 seconds. Thus, the data set consists of 121 single cough records, 48 of which were reported to have a positive PCR test result, and 73 were reported to have a negative test result. The original format of the records (before splitting) is also given in the data set.

### 2.1.2 NoCoCoDa

In [16], public media interviews with COVID-19 patients were manually reviewed and the cough sounds were separated one by one and recorded. These interviews were broadcasted online by news sources. This database, called the NoCoCoDa, contains a total of 73 single cough sounds and is available to researchers free of charge upon request. This data set, with a total of 13 interviews attended by 10 people, includes reflex COVID-19 cough sounds. The cough sounds were recorded as a .WAV file with a sampling frequency of 44.1 kHz. In addition to the data, information about the patients is given in an additional file. Since NoCoCoDa is derived from reports and news programs, other sounds such as speech or music are heard in the background in some cough recordings. In a few, a mixture of throat clearing and coughing was also found. All this is specified in the additional file.

In the present study, these noisy and suspicious coughs were removed from the NoCoCoDa data set and the remaining 59 coughs were used. After combining these two data sets, the distribution of cough sounds between the two classes is given in Table 1. As can be seen, a total of 180 cough sounds from 107 COVID-19 patients and 73 cough sounds from non-COVID-19 patients were used in this study.

**Table 1.** Distribution of cough sounds between the two classes

| Data set | COVID-19 | Non-COVID-19 | Total |
|---|---|---|---|
| **Virufy** | 48 | 73 | 107 |
| **NoCoCoDa** | 59 | 0 | 59 |
| **Total** | 107 | 73 | 180 |

### 2.2. Mel-frequency cepstral coefficients (MFCCs)

MFCCs are one of the popular and successful methods for obtaining features in voice analysis and automatic speech recognition systems [35]. MFCCs is a digital technical analysis that simulates the perception of human ears and is calculated on the basis of Fast Fourier Transform (FFT). Since the characteristics of speech signals remain stable in a very small time interval (about 20-30 ms), they are processed in very short time intervals [36], [37]. This short interval is called the frame. Frames are usually chosen to overlap to make transitions between frames smoother. Similar to the calculation of spectrogram, here, the windowing process takes place to avoid a discontinuity at the beginning and end of the frames. The commonly used window structure is Hamming. After windowing, FFT is applied to transform each frame from the time domain to the frequency domain. The mel unit is a unit designed to imitate the perceptual feature of the human ear. Conversion between the mel scale and the frequency scale is provided by the equation given below.

$$mel\ (f) = 2595 \times \log\left(1 + \frac{f}{700}\right) \quad (1)$$

In this way, MFCCs are the expression of the short-time power spectrum of the sound signal on the mel scale [38], [39]. When MFCCs are calculated for a cough sound, a matrix is obtained in the M×N matrix, where M is the number of MFCCs and N is the number of segments (the number of frames).

In the literature, MFCCs was used for the classification of cough sounds. For example, in [17], the features were extracted by the MFCCs method for the classification of dry and wet coughs. In order to obtain features with the MFCCs method, attention should be paid to important factors, called hyperparameters, which include the type of window used, frame length, frame overlap length, number of segments used for feature extraction, and number of MFCCs. In this study, the chosen window type was Hamming, and the frame overlap length was half of the frame length. The optimum values of the other three hyperparameters (the frame length, number of MFCCs, and number of segments used for feature extraction) were chosen using the LOO-CV strategy.

**2.3. Classification**

Today, classification is used in various fields, from medical or genomic predictions to systems such as spam detection and face recognition, and even in finance [40]. In the classification process, a classifier is trained with samples with certain labels and a model is created. Then, the model is used to guess the label of unknown samples [41]. Many classifiers based on machine/deep learning methods were used in the classification of cough sounds. For example, in [42], Logistic regression (LR), support vector machines (SVM), multilayer perceptrons (MLP), convolutional neural networks (CNN), long-short term memory (LSTM), and residual-based neural network architecture (Resnet50) was used.

In this study, popular classifiers in machine learning systems were used to classify COVID-19 and non-COVID-19 cough sounds. These are SVM, linear discriminant analysis (LDA), k-nearest neighbors (k-NN), and partial least squares regression (PLSR). In SVM, LDA, and k-NN classifiers, two different structures of the model were implemented. In SVM, two different non-linear kernels, namely radial basis function and polynomial kernels were used. In k-NN, Euclidean and Chebyshev distance metrics, and in LDA, linear and quadratic decision surfaces were tested. By adding PLSR to these six classifiers, a total of seven different classifiers were created and the results of each classifier were calculated. To determine the values of the hyperparameters in each classifier, the LOO-CV strategy was used.

## 2.4. Measuring the performance of the system

The performance of a classification system can be measured with different metrics. In this study, the accuracy rate, AUC, F-measure, sensitivity, and specificity were used to measure the performance of the system. There are different strategies for calculating these metrics. One of the popular strategies is LOO-CV [43]. The LOO-CV strategy is adopted in a system when the number of samples in the data set or even just the number of samples in a class is low [44]. In this strategy, the data set containing N samples is divided into two sections. The N-1 sample is used for training the classifier and the single remaining sample is used for testing the model. All the samples are used for testing only once, so the process is repeated N times and, in this way, different metrics can be computed. In this study, the LOO-CV strategy was used to calculate the metrics, taking into account the total number of samples (180 samples) in the two classes.

## 3. Results

A method was proposed based on machine learning to diagnose COVID-19 patients from non-COVID-19 patients by cough sounds. The proposed method was tested on a data set that includes virufy and NoCoCoDa data sets. The features were extracted from cough sounds using the MFCCs method and classified with seven different classifiers. To select values of hyperparameters in feature extraction and classification processes, the accuracy rate metric was calculated according to the LOO-CV strategy. In searching for the optimum value of a hyperparameter, all the other hyperparameters were kept constant. The value reaching the highest accuracy rate in the searched range was selected as the optimum value of that hyperparameter. In all of the steps of the study, for the MFCCs method, the window type was chosen as Hamming, and the frame overlap length was half of the frame length.

## 3.1. Determination of the values of hyperparameters in the feature extraction phase

As mentioned earlier, in the MFCCs method, three hyperparameters were taken into account. These are the frame length, number of MFCCs, and number of segments used for feature extraction. Frames lengths of 512, 1024, 2048, and 4096 samples were tested to determine the optimum frame length. In this case, the number of MFCCs was selected as 13, and the number of segments used for feature extraction was selected as N. That is, for each cough, a 13xN matrix was obtained, and by averaging in all N segments, a 13x1 feature vector was obtained. Then the feature vectors were transferred to the classifiers and classified. The accuracy rate obtained for each frame length based on the LOO-CV strategy is separately presented for each classifier in Table 2. The hyperparameters of the selected classifiers also appear in the table. As can be seen, the Chebychev-kNN classifier successfully classified the cough sounds recorded from COVID-19 and non-COVID-19 patients for the 2048 frame length with an accuracy rate of 0.9389, followed by the Euclidean-kNN classifier with an accuracy of 0.9167. By looking at the results in general, all the classifiers achieved higher accuracy for the 2048 frame lengths. Therefore, for the continuation of the study, the optimum value of the frame length hyperparameter was chosen as 2048 samples.

**Table 2.** Classification results for different frame lengths

| Frame lengths (samples) | Polynomial-SVM Order=2 | RBF-SVM Sigma =1 | Linear-LDA Gamma=0 | Quadratic-LDA Gamma=0 | Chebychev-kNN K=1 | Euclidean-kNN K=1 | PLSR component=13 |
|---|---|---|---|---|---|---|---|
| **512** | 0.8667 | 0.8944 | 0.8944 | 0.8278 | 0.9056 | 0.9111 | 0.8278 |
| **1024** | 0.8722 | 0.8944 | 0.9000 | 0.8389 | 0.9222 | 0.9111 | 0.8444 |
| **2048** | 0.8722 | 0.8944 | 0.9056 | 0.8556 | **0.9389** | 0.9167 | 0.8556 |
| **4096** | 0.8667 | 0.8944 | 0.9000 | 0.8500 | 0.9056 | 0.9167 | 0.8500 |

To determine the optimum number of MFCCs, a scanning between 2 and 39 was performed. For this, as in the previous step, the number of segments used for feature extraction was selected as N. In this way, for example, when the number of MFCCs is 2, two features, and when it is 39, 39 features are extracted. The performance of all the classifiers was measured by the accuracy rate metric using the LOO-CV strategy. The classifiers' hyperparameter was adjusted as in the previous step. The results are given in Fig. 1. As it turns out, Euclidean-kNN achieved the best performance using 19 MFCCs with an accuracy rate of 0.9500 followed by Chebychev-kNN and polynomial-SVM with an accuracy rate of 0.9389. These ratios were obtained using 13 and 17 MFCCs, respectively. Therefore, the number of MFCCs was determined as 19 for the continuation of the study.

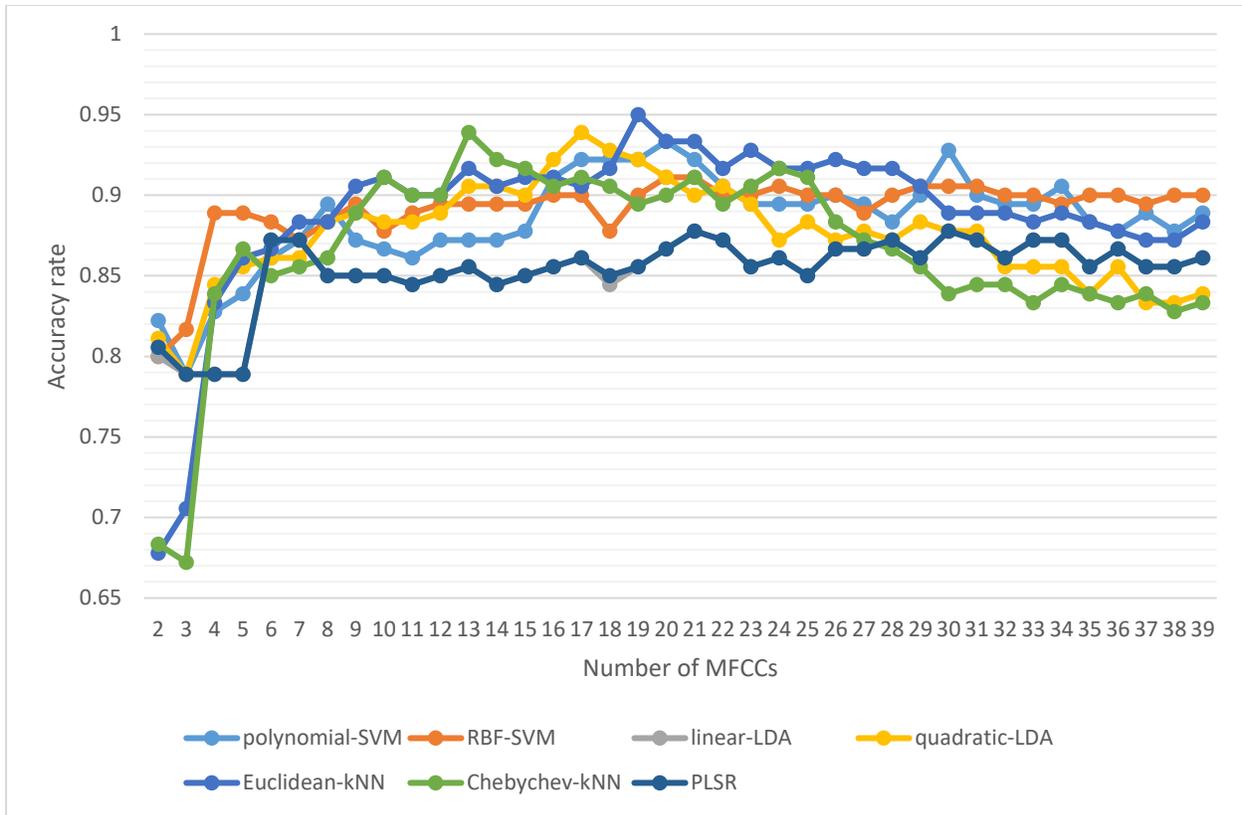

**Fig. 1.** Classification results for different numbers of MFCCs

The last hyperparameter in the feature extraction phase is the number of segments used for this phase. In order to obtain the optimum segment number, numbers from 1 to 50 were used and the averages were obtained. For this purpose, only the first segment of the first 19-MFCCs was used as the feature vector and classified by classifiers. Then, 19 features obtained by the average of the first two segments of 19-MFCCs were classified. This process continued until 50. The hyperparameters of the classifiers were chosen as in the previous steps. The results are shown in Fig. 2. Euclidean-kNN appears to be the most successful classifier in using 17 segments, with an accuracy rate of 0.9833. After determining the optimum values of the hyperparameters in the feature extraction phase, the striking point is the approximately 7% increase in the accuracy of the Euclidean-kNN classifier.

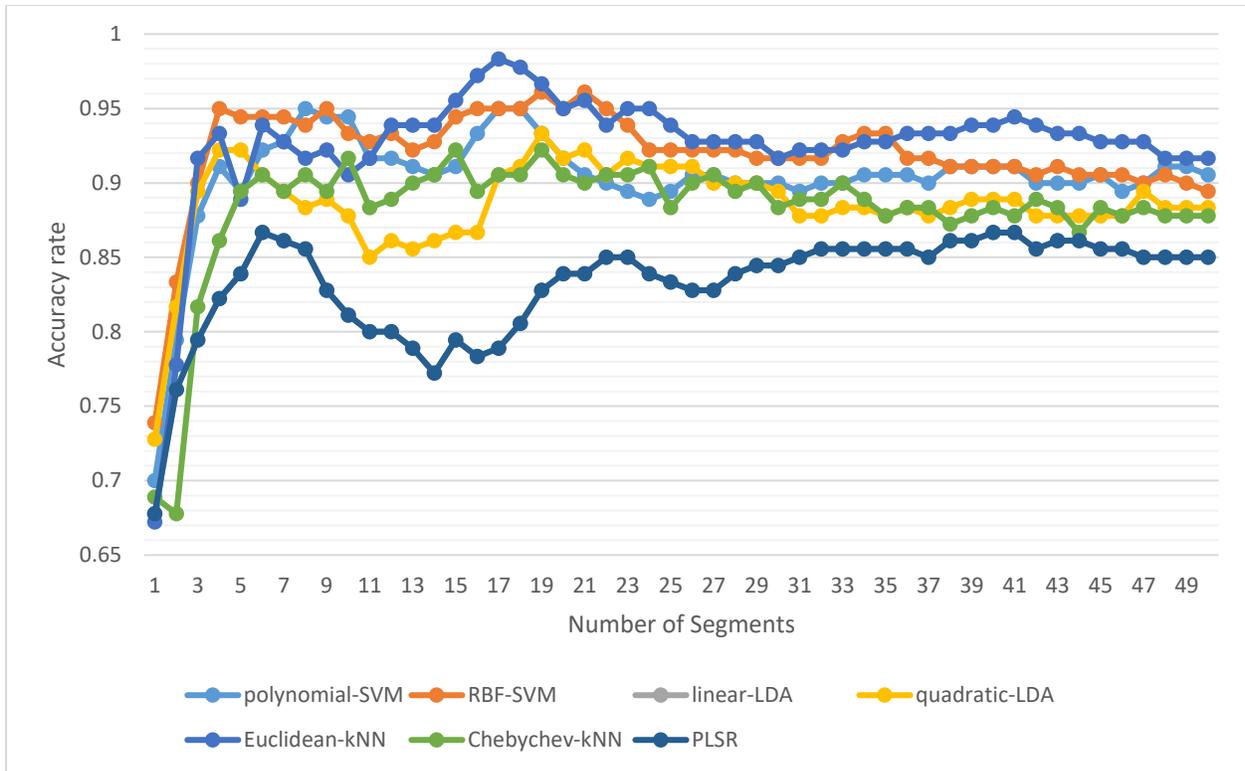

**Fig. 2.** Classification results for different numbers of segments used for feature extraction

### 3.2. Determination of the values of hyperparameters in the classification phase

To determine the optimum hyperparameter values of the classifiers, the accuracy rate metric obtained by the LOO-CV strategy was used. In the RBF-SVM classifier, the range of 0 to 3 was screened by steps 0.1 to determine sigma. The results are given in Fig. A-1. When sigma=1.3, the accuracy rates reached 0.9611. In the polynomial-SVM classifier, the order hyperparameter was searched between 1 to 4. The highest accuracy rate (0.9556) was achieved when the third-order kernel function was used.

Different gammas from 0 to 1 were tested by steps 0.1 to determine gamma in the linear-LDA classifier. The results showed that the highest accuracy was at gamma = 0.6, as given in Fig. A-2. The gamma was changed to 0 and 1 in the quadratic-LDA classifier and the accuracy rate was calculated. A higher accuracy rate (0.9056) was calculated at gamma = 0.

In order to determine k in both types of k-NN classifiers, the classification accuracy rate was calculated from 1 to 25 by steps 1. The results are given in Fig. A-3. The highest accuracy rates at k = 1 were calculated for both classifiers. Thus, the accuracy rates did not change for these classifiers.

Different components were tested by steps 1 from 2 to 19 in order to determine the component in the PLSR classifier. The results show that the highest accuracy rate was when the component = 4, as given in Fig. A-4. Thus, the PLSR classifier classified the cough sounds of COVID-19 and non-COVID-19 patients with an accuracy rate of 0.8111, such as the linear-LDA.

In order to see the system performance more clearly, after determining the optimum hyperparameter values of the classifiers, in addition to the accuracy rate, four more metrics were calculated. The determined hyperparameter values and calculated metrics are given in Table 3. As it turns out, Euclidean-kNN is more successful than the other classifiers in all the metrics. The Euclidean-kNN classifier showed 0.9720 and 1.0000 sensitivity to the COVID-19 and non-COVID-19 class, respectively.

**Table 3.** The results of classification by tuning hyperparameters in classifiers

| Classifier | Hyperparameter | ACC | Sen. Non-COVID-19 | Sen. COVID-19 | F-measure | AUC |
|---|---|---|---|---|---|---|
| Polynomial-SVM | order=3 | 0.9556 | 0.9452 | 0.9626 | 0.9452 | 0.9539 |
| RBF-SVM | sigma=1.3 | 0.9611 | 0.9583 | 0.9630 | 0.9517 | 0.9606 |
| Linear-LDA | gamma=0.6 | 0.8111 | 0.7808 | 0.8318 | 0.7703 | 0.8063 |
| Quadratic-LDA | gamma=0 | 0.9056 | 0.8082 | 0.9720 | 0.8741 | 0.8901 |
| Euclidean-kNN | k=1 | **0.9833** | **1.0000** | **0.9720** | **0.9799** | **0.9860** |
| Chebychev-kNN | k=1 | 0.9056 | 0.8904 | 0.9159 | 0.8844 | 0.9031 |
| PLSR | component =4 | 0.8111 | 0.7808 | 0.8318 | 0.7703 | 0.8063 |

### 3.3. Feature selection based on SFS

In the last step of the study, the feature selection process based on the SFS method was performed separately for each classifier. The results of this step are given in Table 4. In order to see whether the SFS method has any effect, the accuracy rates calculated in the previous step are also shown in the table. As seen, there appears to be an increase for all the other classifiers, with the exception of the Polynomial-SVM and Euclidean-kNN classifiers.

**Table 4.** Effect of the SFS method on classification results

| Classifier | Used Features | ACC (after used SFS) | ACC (before used SFS) |
|---|---|---|---|
| **Polynomial-SVM** | 19 | **0.9556** | **0.9556** |
| **RBF-SVM** | 18 | **0.9667** | 0.9611 |
| **Linear-LDA** | 11 | **0.8388** | 0.8111 |
| **Quadratic-LDA** | 15 | **0.9111** | 0.9056 |
| **Euclidean-kNN** | 19 | **0.9833** | **0.9833** |
| **Chebychev-kNN** | 13 | **0.9444** | 0.9056 |
| **PLSR** | 17 | **0.8277** | 0.8111 |

## 4. Discussion

In this study, for the first time, the cough sounds of conscious and unconscious COVID-19 patients (in the same class) were classified against the cough sounds of non-COVID-19 patients. The results showed that this process was successful by the Euclidean-kNN classifier with an accuracy rate of 0.9833. Thus, it was observed that conscious or unconscious cough sounds did not have any significance in the diagnosis of COVID-19.

As said earlier, few studies have been conducted to distinguish between COVID-19 and non-COVID-19 patients based on cough sounds in the literature. None of these studies have yet been peer-reviewed. However, they are available from preprint banks. Moreover, most of these studies have classified recorded cough sounds that include several (3-5) coughs. In Table 5, the results of these studies and the results of the proposed study are presented for comparison. Except for [33], other studies used data sets that include several cough sounds in each sample. Although samples containing a single cough sound were used in this study, as can be seen, all the metrics of the proposed study were higher than those of the other studies. An important issue to note is the sensitivity of the system in the diagnosis of COVID-19 patients. The proposed study is more successful than other studies with a sensitivity of 0.9720 to COVID-19 patients.

**Table 5**. Comparison of the results of the proposed study with the results of other studies

| Study \ Metrics | ACC | Sen. Non-COVID-19 | Sen. COVID-19 | F-measure | AUC |
|---|---|---|---|---|---|
| [30] | 0.9285 | 0.9114 | 0.9457 | 0.9297 | --- |
| [33] | 0.9586 | 0.9863 | 0.9167 | --- | --- |
| [42] | 0.9501 | 0.9800 | 0.9300 | --- | 0.9632 |
| **Proposed method** | **0.9833** | **1.0000** | **0.9720** | **0.9799** | **0.9860** |

## 5. Conclusion

To diagnosis the COVID-19 disease, it is essential to have a cost-effective, fast, easy, and accurate method, considering the high cost of clinical tests, long turnaround time, and lack of equal access around the world. Therefore, it is quite interesting and essential to distinguish COVID-19 patients from non-COVID-19 ones by evaluating their cough sound via a mobile application based on computer-aided digital technologies. In this way, the user can undergo constant self-surveillance wherever and whenever they want, which leads to infrequent medical visits as well as reducing the crowd in hospitals and the burden of healthcare teams. In this study, a method based on machine learning systems was presented to diagnose COVID-19 and non-COVID-19 patients with a single cough sound. The features were obtained by the MFCCs method from cough sounds and classified with seven different classifiers. The optimum hyperparameters of the system were selected according to the accuracy rate calculated with the LOO-CV strategy. In this way, COVID-19 and non-COVID-19 patients were classified with an accuracy rate of 0.9833, observing an increase in the accuracy of the most successful classifier (Euclidean-kNN) by around 7%. The system showed no error in the diagnosis of non-COVID-19 patients; however, it exhibited a sensitivity of 0.9720 for COVID-19 patients, which shows that it is quite successful compared with the systems available in the literature. Moreover, in this study, the used data set includes the unconscious and reflex cough sounds of COVID-19 patients. The results showed that conscious or unconscious cough sounds did not have any significance in the diagnosis of COVID-19. For future studies, it is planned to test the system on a higher number of samples and on an online platform.

**Acknowledgment**

The author thanks all the healthcare teams for their hard work during the COVID-19 pandemic. Also, the author thanks Madison COHEN-MCFARLANE for sharing the NoCoCoDa data set.

**Appendix**

See Figures A-1, A-2, A-3, and A-4.

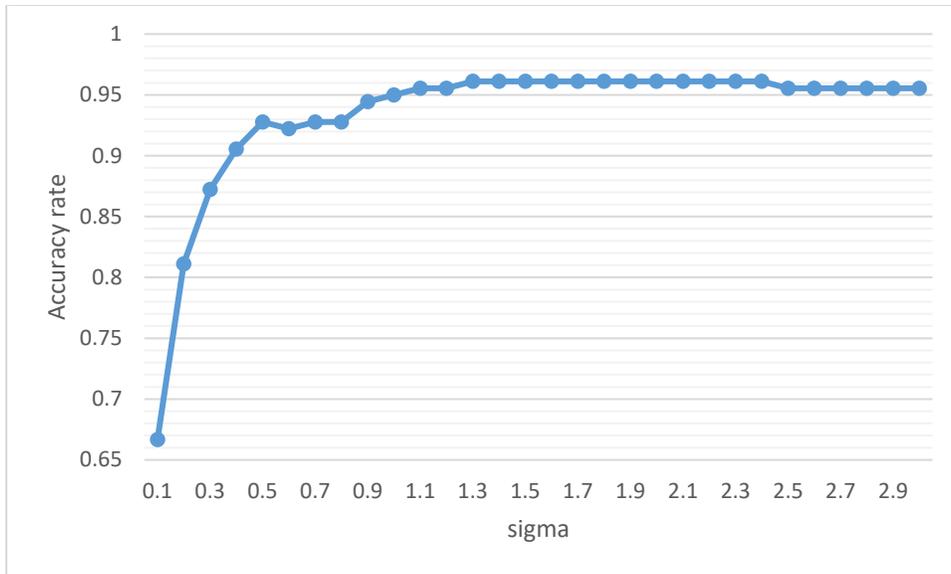

**Fig. A-1.** The accuracy rate of the RBF-SVM classifier for different sigma values

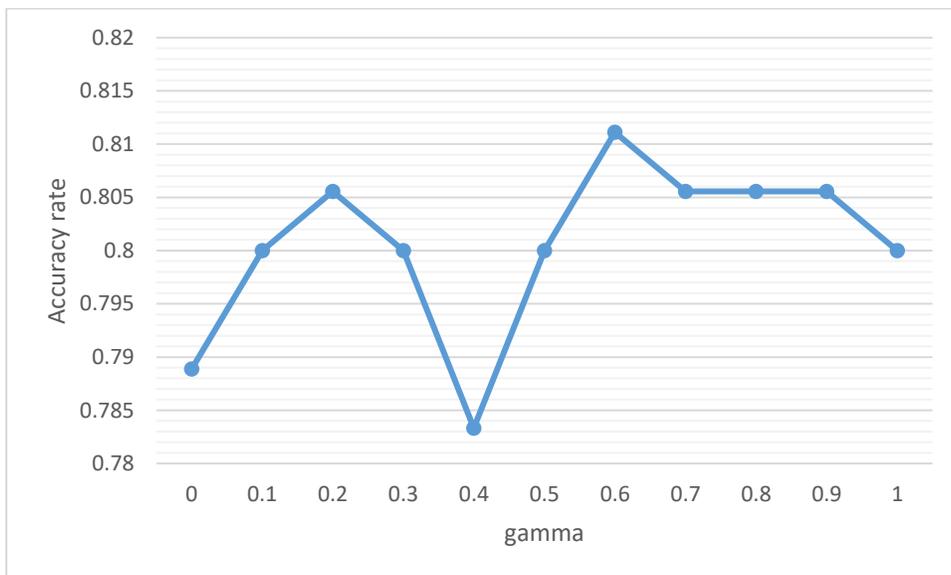

**Fig. A-2.** The accuracy rate of the linear-LDA classifier for different gamma values

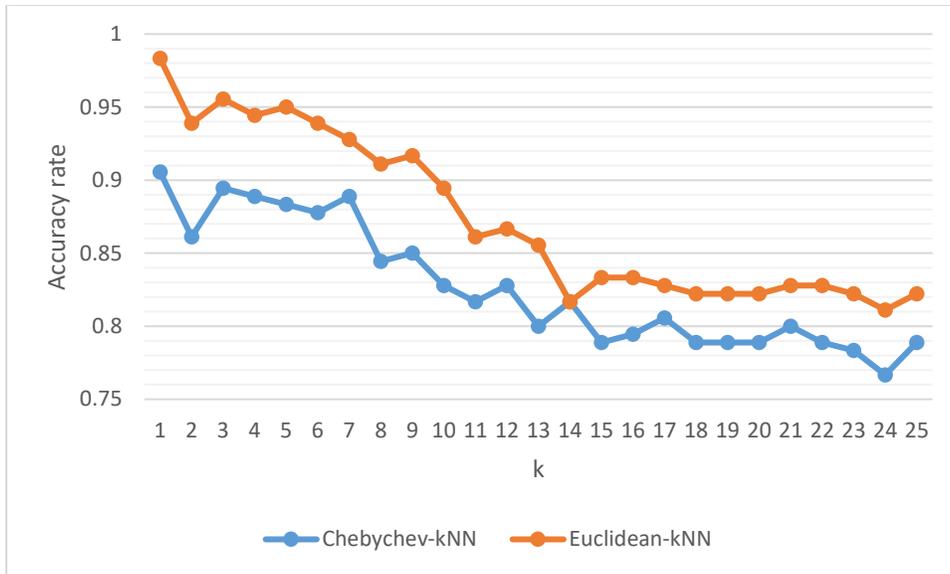

**Fig. A-3.** The accuracy rate of Chebychev-kNN and Euclidean-kNN classifiers for different k values

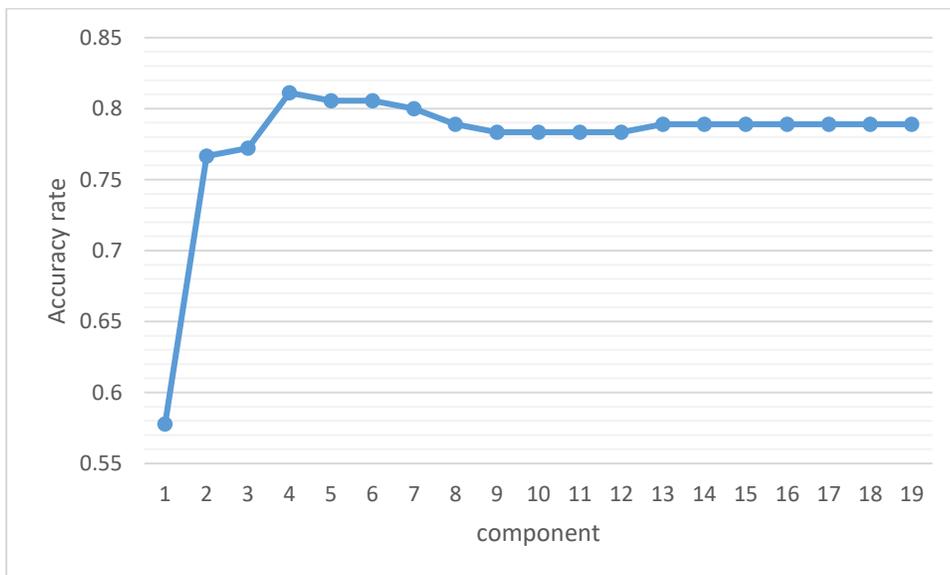

**Fig. A-4.** The accuracy rate of the PLSR classifier for different components